\documentclass[twocolumn,showpacs,aps,prl,superscriptaddress]{revtex4}

\usepackage{graphicx}
\usepackage{dcolumn}
\usepackage{amsmath}
\usepackage{epsfig}

\RequirePackage{xspace}

\usepackage{relsize}
\def\babar{\mbox{\slshape B\kern-0.1em{\smaller A}\kern-0.1em
    B\kern-0.1em{\smaller A\kern-0.2em R}}}

\def\Kbar  {\kern 0.2em\overline{\kern -0.2em K}{}\xspace}

\def\Kz    {\ensuremath{K^0}\xspace}
\def\Kzb   {\ensuremath{\Kbar^0}\xspace}
\def\KzKzb {\ensuremath{\Kz \kern -0.16em \Kzb}\xspace}
\def\Kp    {\ensuremath{K^+}\xspace}
\def\Km    {\ensuremath{K^-}\xspace}

\def\KpKm  {\ensuremath{\Kp \kern -0.16em \Km}\xspace}

\def\Dbar    {\kern 0.2em\overline{\kern -0.2em D}{}\xspace}

\def\Dz      {\ensuremath{D^0}\xspace}
\def\Dzb     {\ensuremath{\Dbar^0}\xspace}
\def\DzDzb   {\ensuremath{\Dz {\kern -0.16em \Dzb}}\xspace}
\def\Dp      {\ensuremath{D^+}\xspace}
\def\Dm      {\ensuremath{D^-}\xspace}

\def\DpDm    {\ensuremath{\Dp {\kern -0.16em \Dm}}\xspace}

\def\Bbar    {\kern 0.18em\overline{\kern -0.18em B}{}\xspace}

\def\BB      {\ensuremath{B\Bbar}\xspace} 
\def\Bz      {\ensuremath{B^0}\xspace}
\def\Bzb     {\ensuremath{\Bbar^0}\xspace}
\def\BzBzb   {\ensuremath{\Bz {\kern -0.16em \Bzb}}\xspace}
\def\Bu      {\ensuremath{B^+}\xspace}
\def\Bub     {\ensuremath{B^-}\xspace}

\def\BpBm    {\ensuremath{\Bu {\kern -0.16em \Bub}}\xspace}

\def\BorBbar    {\kern 0.18em\optbar{\kern -0.18em B}{}\xspace}
\def\DorDbar    {\kern 0.18em\optbar{\kern -0.18em D}{}\xspace}
\def\KorKbar    {\kern 0.18em\optbar{\kern -0.18em K}{}\xspace}

\def\jpsi     {\ensuremath{{J\mskip -3mu/\mskip -2mu\psi\mskip 2mu}}\xspace}

\mathchardef\Upsilon="7107
\def\Y#1S{\ensuremath{\Upsilon{(#1S)}}\xspace}

\def\FourS {\Y4S}

\mathchardef\Deltares="7101
\mathchardef\Xi="7104
\mathchardef\Lambda="7103
\mathchardef\Sigma="7106
\mathchardef\Omega="710A

\def\Deltabar{\kern 0.25em\overline{\kern -0.25em \Deltares}{}\xspace}
\def\Lbar{\kern 0.2em\overline{\kern -0.2em\Lambda\kern 0.05em}\kern-0.05em{}\xspace}
\def\Sigbar{\kern 0.2em\overline{\kern -0.2em \Sigma}{}\xspace}
\def\Xibar{\kern 0.2em\overline{\kern -0.2em \Xi}{}\xspace}
\def\Obar{\kern 0.2em\overline{\kern -0.2em \Omega}{}\xspace}
\def\Nbar{\kern 0.2em\overline{\kern -0.2em N}{}\xspace}
\def\Xb{\kern 0.2em\overline{\kern -0.2em X}{}\xspace}

\def\mes        {\mbox{$m_{\rm ES}$}\xspace}

\def\DeltaE     {\mbox{$\Delta E$}\xspace}

\newcommand{\tev}{\ensuremath{\mathrm{\,Te\kern -0.1em V}}\xspace}
\newcommand{\gev}{\ensuremath{\mathrm{\,Ge\kern -0.1em V}}\xspace}
\newcommand{\mev}{\ensuremath{\mathrm{\,Me\kern -0.1em V}}\xspace}
\newcommand{\kev}{\ensuremath{\mathrm{\,ke\kern -0.1em V}}\xspace}
\newcommand{\ev}{\ensuremath{\mathrm{\,e\kern -0.1em V}}\xspace}
\newcommand{\gevc}{\ensuremath{{\mathrm{\,Ge\kern -0.1em V\!/}c}}\xspace}
\newcommand{\mevc}{\ensuremath{{\mathrm{\,Me\kern -0.1em V\!/}c}}\xspace}
\newcommand{\gevcc}{\ensuremath{{\mathrm{\,Ge\kern -0.1em V\!/}c^2}}\xspace}
\newcommand{\mevcc}{\ensuremath{{\mathrm{\,Me\kern -0.1em V\!/}c^2}}\xspace}

\def\mus  {\ensuremath{\rm \,\mus}\xspace}

\def\mus        {\ensuremath{\,\mu{\rm s}}\xspace}    

\def\pep2{PEP-II}

\def\gsim{{~\raise.15em\hbox{$>$}\kern-.85em
          \lower.35em\hbox{$\sim$}~}\xspace}
\def\lsim{{~\raise.15em\hbox{$<$}\kern-.85em
          \lower.35em\hbox{$\sim$}~}\xspace}

\xspace

\def\jetset74   {\mbox{\tt Jetset \hspace{-0.5em}7.\hspace{-0.2em}4}\xspace}

\def\figurebox#1#2#3{
    \def\arg{#3}
    \ifx\arg\empty
    {\hfill\vbox{\hsize#2\hrule\hbox to #2{\vrule\hfill\vbox to #1{\hsize#2\vfill}\vrule}\hrule}\hfill}
    \else
    {\hfill\epsfbox{#3}\hfill}
    \fi}

\begin{document}


\begin{flushleft}
BaBar-PUB-03/047\\
SLAC-PUB-10332\\
\end{flushleft}

\title{
{\large \boldmath
 Observation of the Decay $B\rightarrow J/\psi \eta K$
and Search for $X(3872)\rightarrow J/\psi \eta$
}
}

%
\author{B.~Aubert}
\author{R.~Barate}
\author{D.~Boutigny}
\author{F.~Couderc}
\author{J.-M.~Gaillard}
\author{A.~Hicheur}
\author{Y.~Karyotakis}
\author{J.~P.~Lees}
\author{V.~Tisserand}
\author{A.~Zghiche}
\affiliation{Laboratoire de Physique des Particules, F-74941 Annecy-le-Vieux, France }
% ...
\collaboration{The \babar\ Collaboration}
\noaffiliation

\date{\today}

\begin{abstract}
We report the observation of the $B$ meson decay 
$B^\pm\rightarrow J/\psi \eta K^\pm$ and evidence for the decay 
$B^0\rightarrow J/\psi \eta K^0_S$, 
using {90} million $\ensuremath{B\Bbar}\xspace$ events collected at the 
$\ensuremath{\Upsilon{(4S)}}\xspace$ resonance with the 
$\mbox{\slshape B\kern-0.1em{\smaller A}\kern-0.1em B\kern-0.1em{\smaller A\kern-0.2em R}}$
detector at the PEP-II $e^+ e^-$ asymmetric-energy storage ring. 
We obtain branching fractions of
$\cal{B}$$(B^\pm\rightarrow J/\psi \eta K^{\pm}$)=$(10.8\pm 2.3(\rm{stat.})\pm 2.4(\rm{syst.}))\times 10^{-5}$ 
and 
$\cal{B}$$(B^0\rightarrow J/\psi\eta K_{\rm{S}}^{0}$)=$(8.4\pm 2.6(\rm{stat.})\pm 2.7(\rm{syst.}))\times 10^{-5}$. 
We search for the new narrow mass state, the $X(3872)$, 
recently reported by the Belle Collaboration, in the decay 
$B^\pm\rightarrow X(3872)K^\pm, X(3872)\rightarrow \jpsi \eta$ and
determine an upper limit of
$\cal{B}$$(B^\pm \rightarrow X(3872) K^\pm \rightarrow \jpsi \eta K^\pm$)
$<7.7\times 10^{-6}$ at 90$\%$ C.L.

\end{abstract}


\maketitle
The study of charmonium states produced in exclusive $B$ meson decays led to
observations of known charmonium states and recently to the discoveries of new
states. Since $B$ mesons can decay via color-suppressed 
$b\longrightarrow c\overline{c}s$ quark transitions, the charmonium states are typically
produced in final states with kaons.
Many known charmonium states have been observed in decays such as 
$B\rightarrow J/\psi K^{(*)}$, $\psi(2S) K^{(*)}$, $\chi _{\rm{c}}K^{(*)}$, and $\eta
_{\rm{c}}(1S)K^{(*)}$ and evidence
for new states  
such as a candidate for the 
$\eta_{\rm{c}}(3654)$ 
has been published~\cite{belle-1}. 
Recently the Belle
Collaboration~\cite{belle-2} observed a new narrow mass state with a $3.872$ $\gevcc$ mass produced in
the decay $B^\pm\rightarrow X\left( 3872\right) K^{\pm },$ $X\left( 3872\right)
\rightarrow \pi ^{+}\pi ^{-}J/\psi $. This new state may be the hitherto
undetected $J^{PC}=2^{--}$ $1^{3}D_{2}$\ \ charmonium state~\cite{quigg}.
However, such a state should have a large radiative $E1$ dipole transition into $\gamma\chi_{\rm{c1}}$,
which Belle does not observe, and theoretical models~\cite{quigg} predict a smaller mass splitting, relative 
to the $\psi (3770)$, than observed. 
Unconventional explanations include a molecule~\cite{tornqvist} formed with charmed $D$ and $
D^{\ast }$ mesons, since the $X(3872)$ has a mass exactly at $D^{\ast 0}\left(
2007\right) +D^0\left( 1864\right) $ threshold. 
Alternatively, 
this new state may be a hybrid
charmonium state~\cite{close} formed of $c\overline{c}+\rm{gluons}$ 
since 
color octet charmonium states may be
produced in exclusive B  
decays~\cite{bodwin-beneke}. 

To further elucidate the nature of the $X(3872)$,
we performed an analysis     
on the new exclusive decay  $B\rightarrow J/\psi \eta K$, 
to search for $X(3872)\rightarrow J/\psi \eta$.
If the $X(3872)$ is a conventional charmonium
state, its decays may be similar to the $\psi(2S), $ which
decays into $J/\psi \pi ^{+}\pi ^{-}$
and, with a factor ten smaller relative rate,
into $\jpsi \eta$.
If instead, it is a
hybrid charmonium state, it is 
also predicted~\cite{close} to decay into $J/\psi\pi \pi$
and $J/\psi \eta$
with possibly an enhanced rate in the $\eta$ channel. 

The decay
$B\rightarrow J/\psi \eta K$ 
is similar at the quark level 
to other color-suppressed decays such as
$B\rightarrow J/\psi \phi K$
which has been observed with a branching fraction of
$(4.4\pm1.4\pm0.5)\times 10^{-5}$~\cite{jpsiphik}. 
Hence it might be expected that $B\rightarrow J/\psi \eta K$ 
has a comparable branching fraction.

The data used in this analysis 
correspond to a total integrated luminosity of 
$81.9$ fb$^{-1}$ taken on the $\FourS$
resonance, producing a sample 
of $90.0\pm1.0$ million $\BB$ events ($N_{\BB
}$).
Data
were collected at
the PEP-II asymmetric-energy $e^{+}e^{-}$ storage ring with the $\babar$ detector, 
fully described elsewhere~\cite{babar-det}. The $\babar$
detector includes a silicon vertex tracker and a 
drift chamber in a 1.5-T solenoidal magnetic field 
to detect charged particles and measure their momenta and
energy loss. Photons, electrons, and neutral hadrons are detected in a
CsI(Tl)-crystal electromagnetic calorimeter. 
An internally reflecting ring-imaging
Cherenkov detector is used
for particle
identification. 
Penetrating
muons and neutral hadrons are identified  
by resistive-plate chambers 
in the steel flux return.
Preliminary 
track-selection criteria in this analysis follow previous
$\babar$ analyses~\cite{babar-charmonium}
and the detailed explanation of the particle identification 
(PID) is given elsewhere~\cite{babar-charmonium},~\cite{kaon-pid}. 

The intermediate states in the 
charged ($J/\psi \eta K^{\pm}$)
and 
neutral ($J/\psi \eta K_{\rm{S}}^{0}$)
modes used in this analysis, 
$J/\psi \rightarrow e^+ e^-$, 
$J/\psi \rightarrow \mu^+  \mu^-$, 
$\eta \rightarrow \gamma\gamma$ 
and $K_{S}^{0}\rightarrow \pi ^{+}\pi ^{-}$, are selected within the mass 
intervals 
$2.95<M(e^{+}e^{-})<3.14$, 
$3.06<M\left( \mu ^{+}\mu ^{-}\right)<3.14$, 
$0.525<M(\gamma\gamma)<0.571$, 
and 
$0.489<M\left( \pi ^{+}\pi^{-}\right)<0.507$ $\gevcc$.
The mass interval for $e^+e^-$ is larger than than for $\mu^+\mu^-$ to enable
detection of events with Bremsstrahlung in the detector.
The $K_{S}^0$ decay length in the lab frame is required to be greater than 0.1 cm. 

Determination of the signal and the background utilizes two kinematic variables~\cite{jpsiphik}: 
the energy difference $\Delta E$ between the energy of the $B$ candidate and 
the beam energy $E_{\rm{b}}^{*}$ in the $\FourS$ rest frame; and the beam-energy-substituted 
mass $\mes =\sqrt{\left( E_{\rm{b}}^{*}\right)^2 -\left( p_{\rm{B}}^{*}\right) ^{2}}$, where 
$ p_{\rm{B}}^{*}$ is the reconstructed momentum of the $B$ candidate in the $\FourS$ frame.
Signal events should be concentrated in a rectangular signal-box region
bounded by 
$|\mes-m_{\rm{B}}| < 7.5$ $\mevcc$, where $m_{\rm{B}}$ is the mass of $B$ meson
and $|\Delta E|<$ 40 MeV. 

Before the data were analyzed, the final selection criteria were optimized 
separately for the charged and neutral modes using 
a Monte Carlo (MC) simulation of the signal and the known backgrounds.
Motivated by the $B\rightarrow J/\psi \phi K$ measurement, the
$ab\ initio$ value of the
branching fraction for $B\rightarrow J/\psi \eta K$ 
used in the signal MC was $5\times 10^{-5}$.
The number of reconstructed MC signal events $n_{\rm{s}}^{\rm{mc}}$ and the number of
reconstructed MC background events $n_{\rm{b}}^{\rm{mc}}$ in the signal-box were used 
to
estimate the sensitivity ratio, 
$n_{\rm{s}}^{\rm{mc}}/\sqrt{n_{\rm{s}}^{\rm{mc}}+n_{\rm{b}}^{\rm{mc}}}$. 
This ratio was maximized by
varying the selection criteria
on  
the $\eta$ mass,
a $\pi^0$ veto, 
the photon helicity angle from
the $\eta$ decay 
and the thrust angle.
The $\gamma\gamma$
mass interval of the $\eta$ candidate as specified earlier 
was chosen by this procedure.
In the charged(neutral) mode,
if either of the photons associated with an $\eta$ candidate,
in combination with any other photon in the event,
forms
a $\gamma \gamma $ mass within 
17(10) $\mevcc$ 
of the nominal $\pi^0$ mass, 
the $\eta$ candidate is vetoed as a $\pi^0$ background.
The $\eta$ candidate is rejected if 
$\left| \cos \theta _{\gamma }^{\eta }\right|$ is greater
than 0.93(0.81), where $
\theta _{\gamma }^{\eta }$ is the photon helicity angle~\cite{babar-charmonium} 
in the $\eta$ rest
frame. 
Signal events 
have a uniform  
$\cos \theta _{\gamma }^{\eta }$ 
distribution
whereas combinatorial background of random pairs of photons
typically has a distribution that
peaks near $\pm 1$.  
\begin{table*} [!htb]
\caption{Efficiencies, number of signal-box and background events, 90\% C.L. of the number of events and the branching fraction upper limits, P-values and branching fractions}
\begin{center}
\footnotesize{
\begin{tabular}{lccccccc}
\hline\hline\\[-0.2cm]

Mode & $\epsilon$ & $n_{0}$ & $n_{b}\pm \sigma _{b}$ & $N_{90\%}$& $90\%$ C.L.U.L.& P-value  & ~~~Branching Fraction \\ 
\hline \\[-0.2cm]

$J/\psi \eta K^{\pm}$ &  $10.75\%$&99 & $50.3\pm 3.0$ &70.0&$< 15.5\times 10^{-5}$&$2\times 10^{-8}$&
$(10.8\pm 2.3\pm 2.4)\times 10^{-5}$ \\ 
$J/\psi \eta K_{S}^{0}$ & ${8.53\%}$&39 & {${18.5\pm 1.7}$} &34.5&$< 14.1\times 10^{-5}$&$9\times 10^{-5}$& 
$(8.4\pm 2.6\pm 2.7)\times 10 ^{-5}$ \\ 
\hline 
\end{tabular}
}
\end{center}
\label{table-upperlimit-bf}
\end{table*}

To separate two-jet continuum events from the
more spherical decays of $B$ mesons produced nearly at rest
from $\FourS\rightarrow  \BB$, the angle $\theta_{\rm{T}}$ between the 
thrust~\cite{babar-charmonium} direction of the $B$
meson candidate and the thrust direction of the remaining charged
tracks and photons in the
event is calculated. We reject events
when  $\left| \cos \theta _{\rm{T}}\right|$ is greater
than 0.8(0.9), 
since 
the distribution 
in $\cos\theta_{\rm{T}}$ 
is flat for $\BB$ events, 
while 
background $e^+e^- \rightarrow q\overline{q}$  continuum events 
peak at $\cos \theta_{\rm{T}}=\pm 1$.

\begin{figure}[!htb]
\begin{center}
\includegraphics[height=7cm]{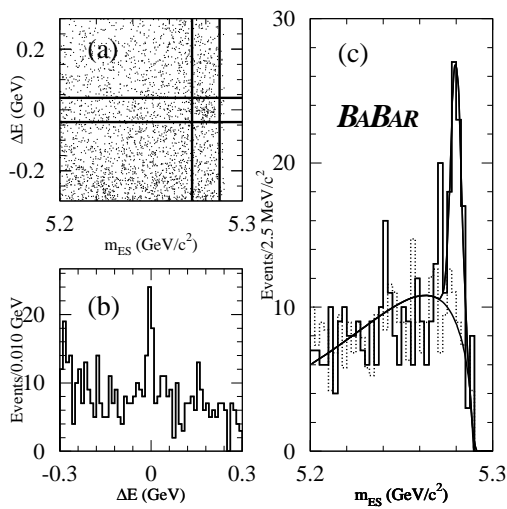}
\caption{ 
For $B^{\pm}\rightarrow J/\psi \eta K^\pm$, 
the $\Delta E$ versus $\mes$ event distribution (a) is shown  
with vertical and horizontal bands defined by limits,  
$|\mes-m_{\rm{B}}|<7.5$ $\mevcc$ 
and $|\Delta E|<$ 40 MeV, respectively.
The intersection of these bands
corresponds to the signal-box region defined
in the text. 
The $\Delta E$ projection (b) is shown  
for events in the vertical band that contains the $\mes$ signal region.
The $\mes$ projection (c) is shown 
for events
in the horizontal band that
contains the $\Delta E$ signal region.
The dashed histogram represents the estimated background and is described in the text.}
\label{fig:jpkub}
\end{center}
\end{figure}

\begin{figure}[!htb]
\begin{center}
\includegraphics[height=7cm]{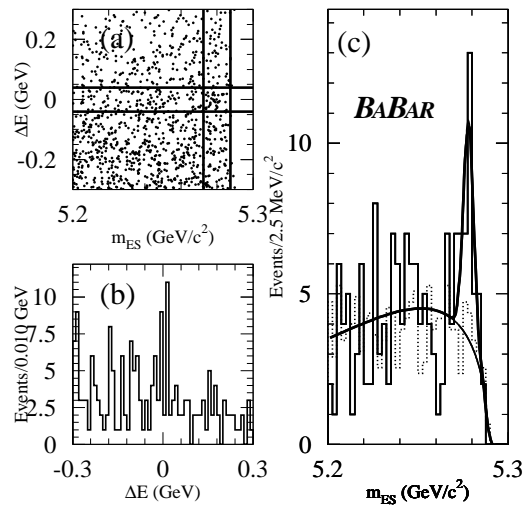}
\caption{The $\Delta E$ and $\mes$ distributions 
for $B^{0}\rightarrow J/\psi \eta K^0_{\rm{S}}$. 
The descriptions of Figs. 2(a), (b), and (c) follow those of Figs. 1(a), (b), and (c), respectively.}
\label{fig:jpksub}
\end{center}
\end{figure}

The data, after these cuts, are 
shown in 
Figs.~\ref{fig:jpkub} and \ref{fig:jpksub} 
where
(a) is a scatter plot of $\Delta E$ versus $\mes$,
(b) is the $\Delta E$ histogram
and
(c) is the $\mes$ histogram (solid line).
We find evidence for $B$ signals in both the 
$J/\psi \eta K^{\pm}$ and $J/\psi \eta K_{\rm{S}}^{0}$ modes.


To determine the
branching fraction for these modes, we first find 
the number of signal events, which is defined as 
$n_{\rm{s}}=n_{\rm{0}}-n_{\rm{b}}$, 
where
$n_{\rm{0}}$ is 
the number of events in the signal-box region,
and
$n_{\rm{b}}$ is 
the estimated number of background events. 
For each mode, 
$n_{\rm{b}}$  
is obtained from fitting the 
$\mes$ distribution    
for events with $|\Delta E|<$ 40 MeV
with
the line shape of a
Gaussian function and an
ARGUS function~\cite{babar-charmonium}, which is an
empirical parameterization of the background shape.
The fit parameters are the normalization and
mean of the Gaussian and the normalization of the
background curve.
The width of the Gaussian is fixed to the value determined by
MC simulation and the shape of the background curve is fixed
to a best fit to the data $\mes$ distribution
with the  $\Delta E$ sideband region 
of $0.10<|\DeltaE|<0.14$ GeV for the $B^\pm$ mode
and 
$0.08<|\DeltaE|<0.28$ GeV for the $B^0$ mode.
Figs.~\ref{fig:jpkub}(c) and \ref{fig:jpksub}(c) 
show
the resulting Gaussian and background
curves (solid) and the background events (dashed histogram) 
from the $\Delta E$ sideband regions
normalized to the data in the signal region. 
Integrating the background curve over the signal-box region we obtain
 $n_{\rm{b}}$
and its uncertainty, $\sigma_{\rm{b}}$.
Results are listed in 
Table~\ref{table-upperlimit-bf}.
Additional checks on the background shapes were performed.
Using data, a $\gamma \gamma$ mass
sideband that is outside the nominal $\eta$ mass was selected and 
a similar $\mes$ background shape was found. 
Using MC simulations of inclusive $B\rightarrow J/\psi$ backgrounds  
another $\mes$ background shape was obtained and fit. 
If we include a fit component due
to this background, which 
is well described by a broad Gaussian distribution,
and we refit the data sideband 
distribution, the background results do not change. 

The branching fraction is calculated as
$\cal{B}$ $=n_\mathrm{s} / (N_{\BB}\times \epsilon \times f)$
where $\epsilon$ is the efficiency and $f$ is the
product of secondary branching fractions
for the $J/\psi ,$ $\eta $, and $K^0_{\rm{S}}$.
Efficiencies  
are determined by MC simulation 
with three-body phase space and
the branching fractions of
$\FourS \rightarrow B^+B^-$ and 
$\FourS \rightarrow \BzBzb$ are assumed to be equal. 
Results on $\cal{B}$ are given 
in the last column of Table~\ref{table-upperlimit-bf} where 
the first and second errors are statistical and
systematic, respectively.
The statistical error is derived from the uncertainty 
in $n_{\rm{s}}$ which is $\sqrt{n_0+\sigma_{\rm{b}}^2}$.

The systematic  
error, $\sigma_{\rm{sys}}$, for each mode (charged /neutral) 
is determined by adding in quadrature
the percentage uncertainty 
on each of the following quantities: 
$N_{\BB}$ (1.1/1.1); 
secondary branching fractions~\cite{pdg} (2.48/2.52); 
MC statistics (1.77/2.17); 
PID, tracking, and photon detection efficiencies (8.2/8.3); 
$\pi^0$ veto (8.1/8.3);
$\eta$ mass range (3.40/3.14);
background parameterization (16.7/27.0);
and
model dependence (5.1/9.5).
The total systematic errors for the
charged and neutral modes are $22.0\%$ and $32.0\%$, respectively.
The uncertainties in the PID, tracking, 
and photon detection efficiencies 
are based on the study of data control 
samples~\cite{babar-charmonium}.
The uncertainty in the $\pi^0$ veto efficiency
was studied by measuring the veto efficiency on 
the inclusive $\eta$ rate 
in data and MC. 
The uncertainty due to the $\eta$ mass selection was determined by
comparing the measured $\eta$ mass resolution in inclusive $\eta$
decays to the $\eta$ mass resolution from the signal MC.
The background parameterization uncertainty was
estimated by changing the ARGUS shape parameter
by $\pm$1 standard deviation, refitting the $\mes$ data distribution,
and recalculating the number of signal events. 
Although this analysis used MC events generated with
three-body phase space 
to determine the final efficiencies,
additional systematic uncertainties due to the decay model dependence are 
estimated. 
The efficiency uncertainty due to
unknown angular distributions and intermediate resonances
has been estimated by comparing the efficiencies obtained in
five different MC generated models. These include 
$100\%$ transversely polarized $J/\psi$,  
$100\%$ longitudinally polarized $J/\psi$, 
large two-body $J/\psi \eta$ mass, 
large two-body $\eta K$ mass
and 
small two-body $J/\psi K$ mass.
The resulting relative change in efficiencies  
was used to estimate
the production model uncertainty.
The resulting total $\sigma_{\rm{sys}}$ for each mode is used to
determine the $\cal{B}$ systematic errors in 
Table~\ref{table-upperlimit-bf}.

The P-value for null hypothesis (no signal) 
is 
the Poisson probability that the background 
events fluctuate to $\ge n_0$. 
Assuming the probability distribution function 
of the background   
is a Gaussian with mean $n_{\rm{b}}$
and 
standard deviation 
$\sigma_{\rm{b}}$, 
we calculate the Poisson probabilities with 
different background values weighted by
this
Gaussian distribution to determine the final P-value
for each mode. The resulting P-values are equivalent to a 
statistical significance of 5.6$\sigma$ and 3.9$\sigma$
for the charged and neutral modes, respectively.

We also determine the 90$\%$ confidence level upper limit (C.L.U.L.) 
on the branching fraction using
 $n_{0}$, $n_{\rm{b}}$, and $\sigma_{\rm{b}}$, 
in the signal region, and $\sigma _{\rm{sys}}$. 
The Bayesian upper limit on the number of signal events, $N_{90\%}$, is obtained by
folding the Poisson distribution with two Gaussian distributions representing  
the background and systematic uncertainties and integrating the resulting function 
to the $90\%$ confidence level (C.L.). 
This assumes that the $a\ priori$  branching fraction distributions are uniform.
The  charged and neutral results, 
$J/\psi \eta K^\pm$ and  $J/\psi \eta
K_{\rm{S}}^{0}$,
are listed in
Table~\ref{table-upperlimit-bf}.

Our resulting branching fractions 
are comparable to  
the color-suppressed decay  
$B\rightarrow J/\psi \phi K$
branching fraction.
The ratio of the charged $(J/\psi \eta K^\pm)$
to neutral $(J/\psi \eta K_{\rm{S}}^{0})$
branching fractions is
consistent within errors to
the expected value of two.

\begin{figure}[!htb]
\begin{center}
\includegraphics[height=3.75cm]{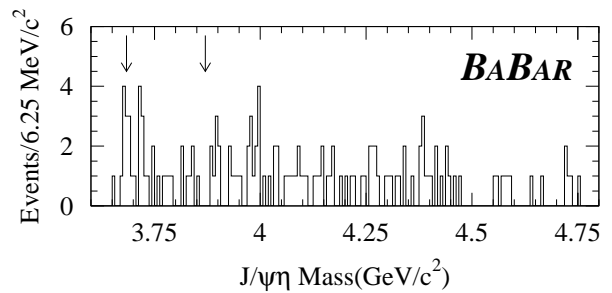}
\caption{The $J/\psi \eta$ mass distributions 
from $B^{\pm}\rightarrow J/\psi \eta K^\pm$ and
$B^{0}\rightarrow J/\psi \eta K^0_{\rm{S}}$.
The arrows indicate where the $\psi(2S)$ and $X(3872)$ signals
would appear.}
\label{fig:jpsieta}
\end{center}
\end{figure}

We search
for the $X(3872)$  
in $B\rightarrow X K, X\rightarrow J/\psi \eta$ 
now only selecting the signal region, 
$|\mes-m_{\rm{B}}|<7.5$ $\mevcc$
and $|\Delta E|<$ 40 MeV. 
The resulting $J/\psi \eta$ 
mass distribution is shown in  
Figure~\ref{fig:jpsieta}.
The two-body
mass resolution from Monte Carlo studies is 6 $\mevcc$. 
There is evidence for the
$\psi(2S)$ and no evidence for the $X(3872)$.
Using the measured branching fraction
$\cal{B}$$(B^\pm\rightarrow \psi(2S) K^\pm)=(6.8\pm0.4)\times 10^{-4}$~\cite{pdg},
we expect to reconstruct $12\pm1$ events in the charged mode in the $\jpsi\eta$ mass
region below 3.710 $\gevcc$ and we observe 15.
After restricting the mass to $3.85<M(\jpsi\eta)<3.89$ $\gevcc$,
we fit the $\mes$ plot with the same procedure as before and
obtain an upper limit for the product branching fraction
$\cal{B}$$(B^\pm \rightarrow X(3872) K^\pm$, $X\rightarrow \jpsi \eta$)
$<7.7\times 10^{-6}$ at 90$\%$ C.L.

Our resulting upper limit 
may be compared to the
Belle result~\cite{belle-2},
${ 
\cal{B} (B^\pm \rightarrow \rm{X}(3872) K^\pm \rightarrow \rm{J}/\psi \pi^+\pi^- K^\pm) 
\over
\cal{B} (B^\pm \rightarrow \psi(\rm{2S}) K^\pm \rightarrow \rm{J}/\psi \pi^+\pi^- K^\pm) 
}$
=
$(6.3\pm 1.2\pm 0.7)\%$. 
Using $\cal{B}$$(B^\pm \rightarrow \psi(2S) K^\pm \rightarrow\jpsi \pi^+\pi^- K^\pm)$
=$(2.0\pm 0.15\pm 0.22)$ $\times$$10^{-4}$~\cite{pdg} 
it can be deduced that
$\cal{B}$$(B^\pm \rightarrow X(3872) K^\pm \rightarrow \jpsi \pi^+\pi^- K^\pm$)=
$(12.6\pm 2.8 \pm 1.2) \times 10^{-6}$.
If the matrix elements for 
$X(3872) \rightarrow \jpsi \pi^+\pi^-$ and $\jpsi \eta$
are similar to those of the $\psi(2S)$ and we include the
larger phase space for the decay of $X(3872)\rightarrow \jpsi \eta$
relative to the $\psi(2S)$, then we would expect
$\cal{B}$$(B^\pm \rightarrow X(3872) K^\pm \rightarrow \jpsi \eta K^\pm$)
$\sim 3\times 10^{-6}$.
Our upper limit is  
within a factor two of this estimate.
This result is not in contradiction
with the charmonium interpretation of the $X(3872)$.

In conclusion, we observe 
the new decay mode
$B\rightarrow$ $J/\psi \eta K$ 
with 
branching fractions of $\cal{B}$$(B^\pm \rightarrow J/\psi \eta K^{\pm}$) =
{${(10.8\pm 2.3\pm 2.4)\times 10}^{-5}$} and 
$\cal{B}$$(B^0\rightarrow J/\psi\eta K_{\rm{S}}^{0}$) = 
{$(8.4\pm 2.6\pm 2.7)\times 10^{-5}$}. 
We set an upper limit for the $X(3872)$
in the product branching fraction,
$\cal{B}$$(B^\pm \rightarrow X(3872) K^\pm \rightarrow \jpsi \eta K^\pm$)
$<7.7\times 10^{-6}$ at 90$\%$ C.L.

\label{sec:Acknowledgments}
We are grateful for the excellent luminosity and machine conditions
provided by our \pep2\ colleagues, 
and for the substantial dedicated effort from
the computing organizations that support \babar.
The collaborating institutions wish to thank 
SLAC for its support and kind hospitality. 
This work is supported by
DOE
and NSF (USA),
NSERC (Canada),
IHEP (China),
CEA and
CNRS-IN2P3
(France),
BMBF and DFG
(Germany),
INFN (Italy),
FOM (The Netherlands),
NFR (Norway),
MIST (Russia), and
PPARC (United Kingdom). 
Individuals have received support from the 
A.~P.~Sloan Foundation, 
Research Corporation,
and Alexander von Humboldt Foundation.

\end{document}